# Mixed Valence Pseudobrookite $Al_{1.75}Ti_{1.25}O_5$: High Temperature Phase Transitions, Magnetism and Resistivity

Davor Tolj[1], WenHua Bi[2], Yong Liu[2], Ivica Zivkovic[1], Henrik M. Ronnow[1], Arnaud Magrez[2,*]

[1]Laboratory of Quantum Magnetism, [2]Crystal Growth Facility

Institute of Physics, Ecole Polytechnique Fédérale de Lausanne - EPFL, Switzerland

*Corresponding author: arnaud.magrez@epfl.ch

## Abstract

Dark blue single crystals of $Al^{3+}_{1.75}Ti^{4+}_{1.0}Ti^{3+}_{0.25}O_5$ were grown with a novel synthesis method based on the reaction of a $Ti^{3+}/Ti^{4+}$ containing langbeinite melt and $Al_2O_3$. The obtained needles crystallize in the pseudobrookite structure and undergo two reversible phase transitions from orthorhombic Cmcm to C2/m first and subsequently to C2 symmetry. Like the known aluminum titanate pseudobrookites, anistropic thermal expansion is observed. The temperature evolution of the crystal structure reveals some insights into the mechanism leading to the decomposition of the $Al_{1.75}Ti_{1.25}O_5$ above 725°C. The magnetic and electrical properties are discussed and compared to other reported aluminum titanate pseudobrookites.

## Introduction

Pseudobrookite was discovered in 1878 on the Uroi Hill in Romania by A. Koch.[1] The appearance of the crystals, the crystal system as well as the physical and chemical properties are very reminiscent of brookite, the orthorhombic polytype of $TiO_2$. By closely examining the shape and measuring the main angles between the facets of the crystals discovered by A. Koch, Prof von Rath proved that the crystals were a "false" brookite and gave the name of pseudobrookite to the new mineral.[2] In 1930, L. Pauling determined the atomic arrangement of pseudobrookite by X-ray diffraction and confirmed that there were no structural relationships between brookite and pseudobrookite crystals.[3][4] The later have a formula $Fe_2TiO_5$ and crystallize in the Cmcm space group (when the lattice parameter *a* is the smallest axis and *c* the largest). In the original description given by L. Pauling, the structure is composed of oxygen octahedra containing iron or titanium. Each iron-containing octahedron shares one edge with another iron-containing octahedron and three edges with titanium-containing octahedra, while each titanium-containing octahedron shares six edges with iron-containing octahedra. The two octahedra are strongly distorted, and are arranged to form c-oriented double chains weakly bonded by the shared edges. The framework of the structure gives rise to rhombus-shaped open channels that extend along the c-axis (Fig. 1a). It was shown in synthetic and natural pseudobrookites that Fe and Ti can substitute for each other at both metal sites, so that the solid solution between $Fe_2TiO_5$ and $FeTi_2O_5$ has varying proportions of $Fe^{2+}$ and $Fe^{3+}$.[5][6]

Although the most studied pseudobrookites have a composition $M_2^{3+}Ti^{4+}O_5$ (M= Sc, Cr, Fe, Ti, Ga, Al), or $M^{2+}Ti_2^{4+}O_5$ (M = Mg, Fe, Co), isomorphic niobates, tantalates, zirconates, antimonates, vanadates, as well as nitrides, oxynitrides and rare earth tetraoxybromides, have been synthesized.[7][8][9][10][11][12]

Such a broad compositional spectrum allows pseudobrookite to be found in a wide variety of applications. For instance, $Ti_{3-d}O_{4-x}N_x$, formed by the N substitution of the high temperature phase of anosovite α-$Ti_3O_5$, isostructural of pseudobrookite, has a band gap of 2.6eV and exhibits superior photocatalytic performance to $TiO_2$.[12] $Ta_3N_5$ materials, which exhibit one of the highest solar to hydrogen conversion efficiencies, are among the most effective compounds for renewable hydrogen production via water splitting.[13] The topotactic insertion of the alkali metals in the open channels gives pseudobrookite materials the potential to be applied in Li-ion batteries.[14][15] Pseudobrookites with lattices containing magnetic elements, in which the exchange interactions are frustrated by the site symmetry of the moments, are of great fundamental interest. This leads to a plethora of physical phenomena that can occur in such geometrically frustrated systems. $CoTi_2O_5$ and $FeTi_2O_5$ are two examples of such systems, both of which exhibit a long-range antiferromagnetic ordered state with a spin-driven Jahn-Teller lattice distortion.[9][10]

Aluminum titanates ($Al_{2-x}Ti_{1+x}O_5$) are the most applied pseudobrookites. With high aluminum content ($Al_2TiO_5$), they are widely used in high-temperature applications where thermal shock resistance and thermal insulation is required, such as the thermal pigments and barriers, internal combustion engine components and metallurgy.[16][17][8][18] The strong anisotropy of thermal expansion generates localized internal stress and causes severe microcracking in pseudobrookites. However, addition of appropriate stabilizers such as $SiO_2$, $Fe_2O_3$ and MgO enhances the mechanical strength performance and improves the sinterability of the ceramics.[19] $Al_2TiO_5$ was also used as a precursor to produce AlTi alloys for aerospace industry by magnesiothermic reduction.[20] Skala et al. have shown that aluminum stoichiometry can be higher than 2 in aluminum titanate pseudobrookites at high temperature.[17] $Al_{2-x}Ti_{1+x}O_5$ pseudobrookites with $0 < x < 1$ are less studied although $Al_2TiO_5$ and $AlTi_2O_5$ crystallize in the same orthorhombic Cmcm structure. The solid solution seems to be discontinued and the substitution of $Al^{3+}$ by $Ti^{3+}$ was found to be limited to $x \leq 0.4$ in polycrystalline samples.[21] Aluminum titanate pseudobrookite with high titanium content ($1 \leq x \leq 2$) shows a rich structural phase diagram with three different structures at room temperature including a distorted monoclinic pseudobrookite structure. At room temperature, $Ti_3O_5$ (x=2) exhibits a different monoclinic structure that transits to the orthorhombic pseudobrookite structure at about 500K. The presence of $Ti^{3+}$ ($3d^1$, spin 1/2) and $Ti^{4+}$ ($3d^0$, spin 0) in these compounds has led to an intensive characterization of their magnetic and transport properties which are dependent on x. While $Al_{2-x}Ti_{1+x}O_5$ pseudobrookites behave as an isolated spin system when $x \leq 1.5$, non-magnetic $Ti^{3+}$-$Ti^{3+}$ dimers develop when Ti-Ti distance shortens with increasing x.[22] A Charge Density Wave (CDW) is proposed theoretically in $AlTi_2O_5$ (x=1) but the material experimentally shows no sign of a CDW state.[22][23] All $Al_{2-x}Ti_{1+x}O_5$ pseudobrookites show an insulating behavior with a band gap narrowing from 0.1eV to 0.05eV when x is increased from 1 to 2.[23]

Herein, we report a facile growth process to produce millimeter-sized single crystals of $Al_{1.75}Ti_{1.25}O_5$. The crystals undergo phase transitions at 550°C and at 650°C. The thermal expansion behavior, as well as the magnetic and transport properties of $Al_{1.75}Ti_{1.25}O_5$ are also discussed.

## Experimental details

### Synthesis

Single crystals of $Al_{1.75}Ti_{1.25}O_5$ were grown using single phase $K_{1.7}Ti_2(PO_4)_3$ powder (prepared similarly as described elsewhere).[24] The powder was thoroughly ground, placed in an α-alumina crucible and heated at 1550 °C for 24h in a pure argon atmosphere. Melt was then cooled down to 1000 °C at a rate of 3°C/h. Subsequently, the heating was stopped and furnace cooled down to room temperature naturally.

**SEM-EDX and XRF**

Sample morphology and composition information were observed by scanning electron microscopy (SEM, The Gemini 300 with Oxford Inst. EDX detector). Energy-dispersive X-ray spectroscopy (EDX) analysis was performed on multiple single crystal samples from the batch to get precise elemental ratio information. Aluminum and titanium ratio were further confirmed using X-ray fluorescence (Orbis PC Micro EDXRF analyzer).

**X-Ray diffraction**

Powder XRD

Precursor, powder and single crystals of $Al_{1.75}Ti_{1.25}O_5$ were characterized by powder X-ray diffraction (PXRD) at room temperature on a Malvern-Panalytical diffractometer (Empyrean system) with Cu K$\alpha$ radiation ($\lambda$ = 1.5148 Å) operating at 45 kV and 40 mA in Bragg-Brentano geometry. Patterns were collected between 10 and 100° in 2θ with step size of 0.013°).

Single crystal XRD

A high quality crystal with suitable size was selected and mounted on a goniometer head with a cryo-loop. Frames were collected at 100K, 200K and 300K on a Rigaku Synergy-I XtaLAB Xray diffractometer, equipped with a Mo micro-focusing source ($\lambda K\alpha$ = 0.71073 Å) and a HyPix-3000 Hybrid Pixel Array detector (Bantam). The temperature was controlled by a Cryostream 800 from Oxford Cryosystems Ltd. CrysAlis[Pro],[25] and OLEX$_2$ software,[26] were used for data reduction and structural refinements, respectively. Structure solutions were obtained with ShelXT program.[27] All other experimental details are listed in Table S1. Structures at 300K, 200K and 100K are available as CIF files with CSD numbers 2221313, 2221314, and 2221315, respectively. DIAMOND program from Crystal Impact was used for crystal structure plotting.[28]

High Temperature Powder XRD

Few needles of $Al_{1.75}Ti_{1.25}O_5$ were ground into fine powder. It was subsequently sealed under vacuum in a quartz capillary which was mounted in the high temperature chamber installed on the powder X-ray diffractometer. Patterns were recorded between 25°C and 1000°C at every 25°C. Rietveld refinement were performed using the Fullprof package.[29]

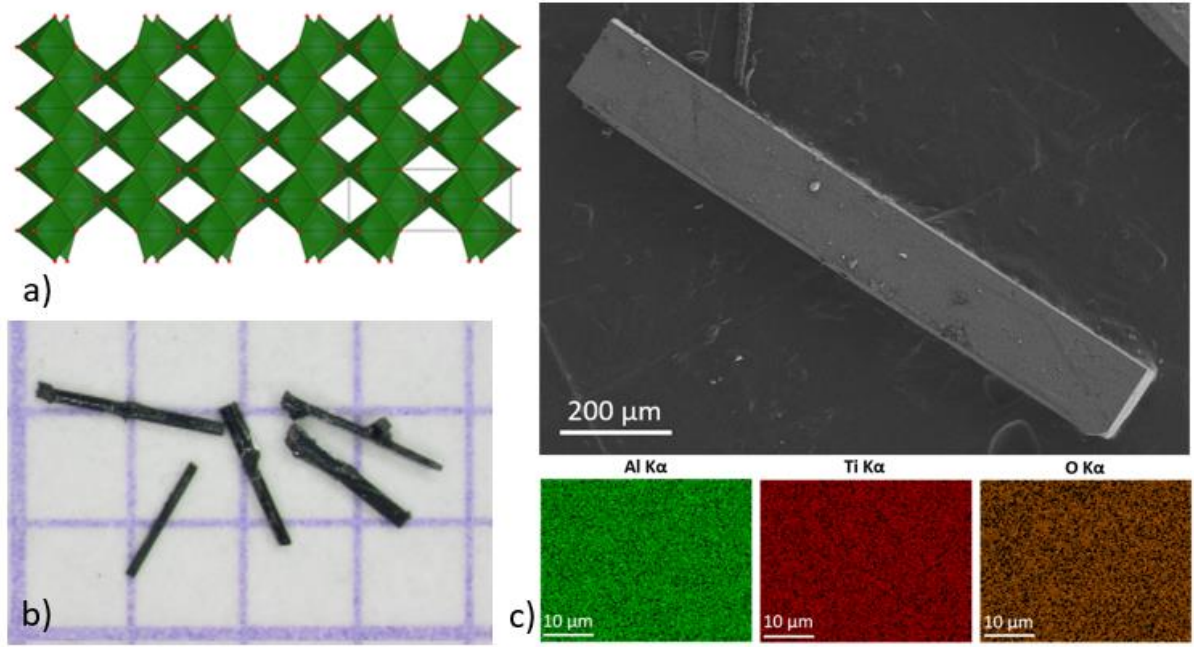

**Fig. 1** a) The 3D packing along c axis. b) Optical photograph of $Al_{1.75}Ti_{1.25}O_5$ single crystals on a millimeter scale paper. c) SEM image of a single crystal (upper panel) with EDX elemental maps (bottom panels).

## Results and discussion

### Growth of single crystals

To the best of our knowledge, investigations of $Al_{2-x}Ti_{1+x}O_5$ have been limited to polycrystalline samples with the exception of aluminum titanate pseudobrookites with $1 \leq x \leq 2$ studied as single crystals grown by the floating zone (FZ) method. For the FZ growth, pseudobrookites were first prepared by a solid-state reaction between $Al_2O_3$, $TiO_2$ and $Ti_2O_3$ at 1000°C in air. The resulting polycrystalline samples were then pressed into feed and seed rods for the FZ process. Although no information on the growth conditions is given in the paper other than the growth atmosphere being a mixture of Ar and $H_2$,[30] the temperature required to melt the rods and stabilize the molten zone should be greater than 1850°C, according to the $Al_2O_3$-$TiO_2$ phase diagram proposed by D. Goldberg.[31] On the other hand $K_{1.7}Ti_2(PO_4)_3$ is a mixed valence (1.3 $Ti^{4+}$ + 0.7 $Ti^{3+}$, average oxidation state of Ti is 3.6) langbeinite that melts at about 1250°C.[32] It is well-known that molten phosphates react strongly with alumina crucibles. Consequently, other crucibles like platinum are preferred for the growth of phosphate single crystals.[33] We took advantage of the known alumina diffusion in molten phosphates to grow $Al_{1.75}Ti_{1.25}O_5$ single crystals in $K_{1.7}Ti_2(PO_4)_3$ flux using the alumina crucible as aluminum source. After the growth, dark blue needle-like crystals were found attached to the alumina crucible. They can easily be removed mechanically from the crucible. The typical size of the needles is about 1.5×0.2×0.15 mm (Fig. 1b). On the other hand, reaction of langbeinite precursor with alumina powder under the same conditions (instead of direct reaction with alumina crucible) didn't produce single crystals but rather resulted in a polycrystalline product with the same composition. The compositional SEM-EDX analysis of the singe crystals showed an average element ratio of Al:Ti:O = 1.75:1.25:5. This composition was confirmed by XRF using $Al_2O_3$/$TiO_2$ standards for accurate quantification of aluminum and titanium stoichiometry. Chemical mappings as well as line scans reveal the Al and Ti to be homogeneously distributed (Fig. 1c) over the needles. In total, more than 5 needles on multiple positions were measured by either SEM-EDX or XRF confirming the homogeneity of the chemical composition between needles. No foreign elements like K or P are found in the needles.

The composition was further confirmed by structure refinement. The obtained site occupancies for Al and Ti give a stoichiometry of Al:Ti:O = 1.758(7):1.242(7):5 after refinement which is in excellent agreement with the SEM-EDX and XRF results. The crystallographic information can be found in the Table 1 in supplementary information. In $Al_{1.75}Ti_{1.25}O_5$, the average oxidation state of titanium is 3.8 similar to the one in the langbeinite precursor. The use of an argon atmosphere limits the oxidation of Ti during the crystal growth. The chemical formula of the pseudobrookite needles can therefore be written $Al^{3+}_{1.75}Ti^{4+}_{1.0}Ti^{3+}_{0.25}O_5$.

**Structure refinement**

The crystal structure was determined based on a single crystal diffraction data collected at 300 K. As shown in Fig. S1, the reconstructed reciprocal space is consistent with the orthorhombic Cmcm space group (100% of the diffraction spot indexed). The symmetry, the refined lattice constants and the atomic coordinates confirm the structure of the needles to be pseudobrookite. In Fig. 2, the room temperature lattice parameters of $Al_{1.75}Ti_{1.25}O_5$, obtained by Rietveld refinement, are compared to the lattice parameters of the solid solution $Al_{2-x}Ti_{1+x}O_5$ versus the oxidation state of Ti.[34] As expected, the lattice parameters evolve linearly with x and consequently with the average oxidation state of Ti. The charge neutrality and the oxygen stoichiometry of pseudobrookites impose that for every Al substituted by Ti, additional $Ti^{3+}$ needs to be present. Therefore, the average oxidation state of Ti in $AlTi_2O_5$ is 3.5+ while $Al_2TiO_5$ only contains $Ti^{4+}$.

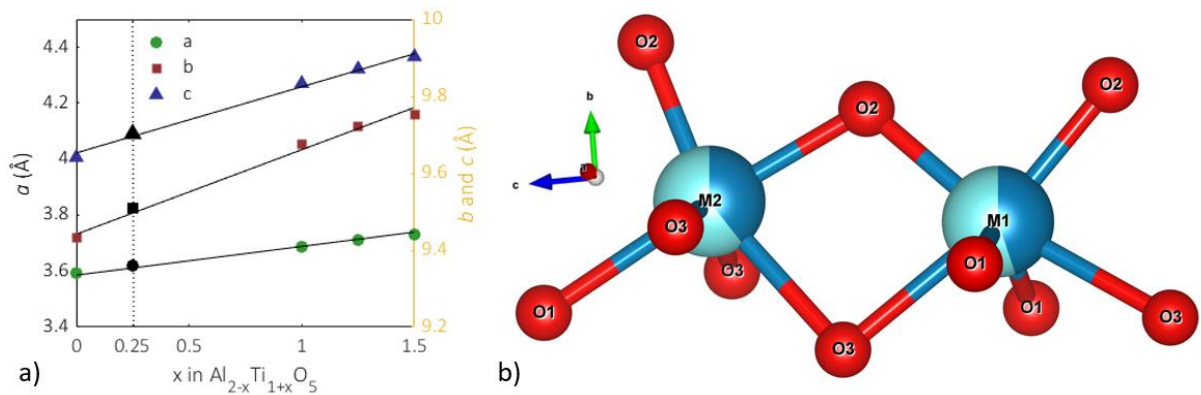

**Fig. 2** a) Lattice parameters at room temperature obtained via Rietveld analysis (black markers at x = 0.25 correspond to the title compound parameters) in dependence of x in $Al_{2-x}Ti_{1+x}O_5$. b) M1-M2 octahedrons bonding motif (Al/Ti ratio depicted in light/dark blue, respectively).

As shown in Fig. 2b, both M1 and M2 containing octahedra in the pseudobrookite structure are strongly distorted. The M1 containing octahedron, with C2v symmetry, can be described with two M-O interatomic distances (i.e. 1.868 Å and 2.107 Å) and three O-M-O angles (80.28, 88.12 and 111.32°). The octahedron interatomic distances increase with the Ti content since the ionic radius of $Al^{3+}$ is smaller than the ones of $Ti^{3+}$ and $Ti^{4+}$. However, the equatorial plane angles remain very similar to those in $Al_2TiO_5$ and $AlTi_2O_5$ structures.[17][23] The M2 containing octahedron has a C1v symmetry. Therefore, the four equatorial interatomic M-O distances are different. They range from 1.831 to 2.139 Å while the four O-M-O angles varies between 82.16 and 101.32°. Both M1 and M2 containing octahedra are described with one apical M-O distance as the equatorial plans are parallel to a mirror plane. In $Al_2TiO_5$,[17] $Al_{1.75}Ti_{1.25}O_5$ and $AlTi_2O_5$, [23] the apical distance of the M2 containing octahedron is larger than in the M1 containing octahedron. The distances behave versus the Ti content similarly to the equatorial M-O distances. On the other hand, no clear trend can be seen in the evolution of the O-M-O apical angle versus the titanium content in both M1 and M2 containing octahedra.

Low temperature SC diffraction data shows that the $a_{orth}$-lattice parameter of the $Al_{1.75}Ti_{1.25}O_5$ contracts from 100K to 300K with a coefficient of thermal expansion (CTE) of -8.92 $10^{-6}$ $K^{-1}$ while $b_{orth}$- and $c_{orth}$- lattice parameters expand with close CTE of 4.08 $10^{-6}$ $K^{-1}$ and 3.90 $10^{-6}$ $K^{-1}$ respectively. Similar thermal expansion anisotropy has been observed in other aluminum titanate pseudobrookites.[35]

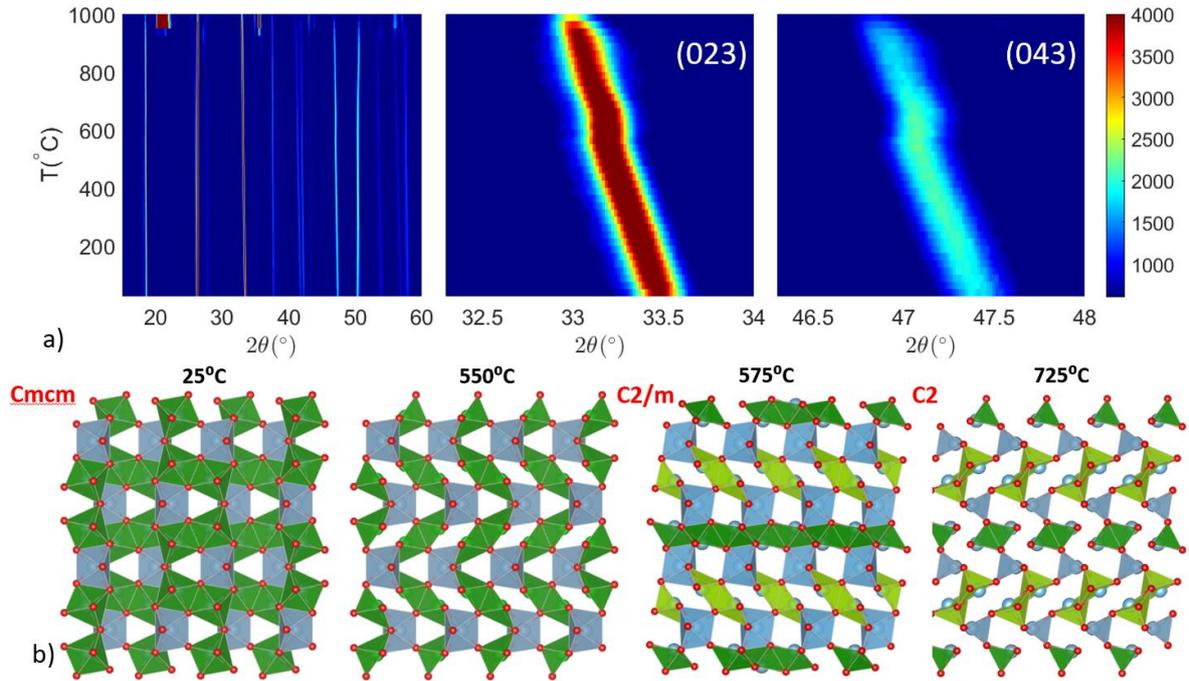

**Fig. 3** a) Contour plot showing the evolution of the XRD pattern with a temperature. b) 3D packing view of $Al_{1.75}Ti_{1.25}O_5$ structure at different temperatures (Orthorombic structure - M1/M2 containing polyhedrons in blue/green, respectively; Monoclinic structure – M1/M2a/M2b containing polyhedrons in blue/dark green/light green, respectively).

**High temperature behavior**

As can be seen in Fig. 3, the temperature evolution of the powder patterns at high temperature clearly shows discontinuities at 550°C as well as at 650°C (Fig. 3a), and are best seen with the shift of 0kl peaks position with temperature. Above 725°C, additional XRD peaks are assigned to rutile and corundum. Their appearance and their increasing height with temperature illustrate the thermal decomposition of $Al_{1.75}Ti_{1.25}O_5$ as previously observed in all aluminum titanate pseudobrookites. [35] Above 950°C, the intensity of pseudobrookite peaks is decreasing quickly, while the one of rutile and corundum peaks keep increasing. The quartz capillary starts crystallizing and strong $SiO_2$ peaks can be observed.

From room temperature to 550°C, the lattice parameters of $Al_{1.75}Ti_{1.25}O_5$ were refined in the Cmcm orthorhombic symmetry. Although the absolute values of CTE measured at low temperature from single crystal XRD data cannot be directly compared with those measured at high temperature from powder XRD data, the thermal evolution of the lattice parameters determined below and above room temperature are consistent in their trend. While CTE along $b_{orth}$- and $c_{orth}$-axis are constant ($1.029 \; 10^{-5}$ $K^{-1}$ and $1.895 \; 10^{-5}$ $K^{-1}$, respectively) over this temperature range as indicated by the linear evolution of the lattice parameters versus temperature, the CTE along the a-axis is temperature dependent. As can be seen in Fig. 4a, the $a_{orth}$-lattice parameter contracts up to about 300°C (CTE= -2.843 $10^{-5}$) while it is almost temperature independent between 300°C and 550°C (CTE= -5.30 $10^{-7}$).

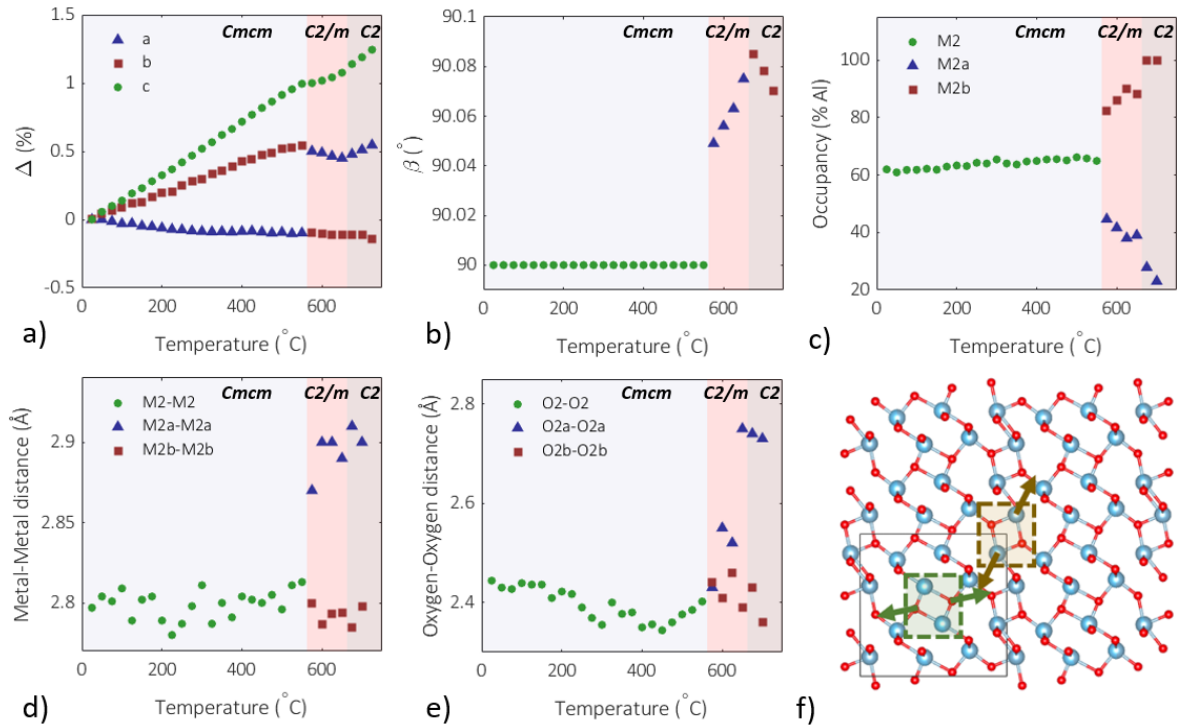

**Fig. 4** Temperature dependence of a change in: a) lattice parameters, b) beta angle, c) M2 site occupancy, d) M-M distances, e) O-O distances. f) Expansions in M2b-O2a-M2b-O2a (green square) and M2a-O2b-M2a-O2b (red square) units in monoclinic structure.

Between 550 and 575°C, $Al_{1.75}Ti_{1.25}O_5$ undergoes a first phase transition from an orthorhombic Cmcm symmetry to a C2/m monoclinic symmetry. A similar transition was already observed with the increase of Ti content in the solid solution $Al_{2-x}Ti_{1+x}O_5$ (1.5 < x ≤ 2, β and λ-phases).[22] As can be seen in the Fig. 4b, the $β_{mono}$ angle increases almost linearly in the 575°C – 675°C temperature range. The $b_{mono}$- corresponding to the $a_{orth}$-lattice parameter remains almost temperature independent (CTE = -2.25 $10^{-6}$ $K^{-1}$) while the CTE of the $c_{mono}$- corresponding to the $c_{orth}$-lattice parameters is reduced to 9.97 $10^{-6}$ $K^{-1}$ as compared to the CTE along the same axis in the 25°C – 550°C temperature range. On the other hand, the $a_{mono}$- corresponding to the $b_{orth}$-lattice parameter starts contracting from 550°C to 675°C with a CTE= -7.04 $10^{-6}$ $K^{-1}$. A second discontinuity in the temperature evolution of the lattice parameters (Fig. 4a) as well as of the peak positions (Fig. 3a) is observed between 675°C and 700°C. It corresponds to a second phase transition from the C2/m to a C2 monoclinic structure. From 700°C to the temperature of thermal decomposition of $Al_{1.75}Ti_{1.25}O_5$, the β angle decreases, the $b_{mono}$ shrinks (CTE= -9.36 $10^{-6}$ $K^{-1}$) while $a_{mono}$ and $c_{mono}$ expands similarly with a CTE of 1.560 $10^{-5}$ $K^{-1}$ and 1.673 10$^{-5}$ K$^{-1}$ respectively. These CTE are very close to the ones measured below 550°C when Al$_{1.75}$Ti$_{1.25}$O$_5$ exhibits the Cmcm symmetry.

We performed a Rietveld refinement of the Al$_{1.75}$Ti$_{1.25}$O$_5$ structure with each powder pattern from room temperature to 725°C in order to study the evolution of the interatomic distances, the coordination as well as the Al- and Ti- distribution in the lattice (site occupancy) leading to the thermal decomposition of Al$_{1.75}$Ti$_{1.25}$O$_5$ (Fig. 4).

Interatomic distances

The shortest distance between sites occupied by M and between sites occupied by O atoms can be found in the center of the lattice. At room temperature, the M2–M2 distance is 2.797 Å. The site occupancy is 62% by Al (i.e. 38% by Ti), Fig. 4c. Considering the size and stoichiometry of Al$^{3+}$, Ti$^{3+}$ and Ti$^{4+}$, the average ionic radius of the atom occupying the M2 site is 0.567 Å. As temperature rises to 550°C, the average ionic radius of the M2 is decreasing following an increase of the site occupancy by Al. Simultaneously, the M2–M2 distance is constant (Fig. 4d). For the same temperature range, the distance M1–M1 as well as M1–M2 are approximately 3.08 Å and 3.61 Å, respectively. They remain almost unaffected by the raise of the temperature and the increase of the average ionic radius of the atom located in the sites.

As refined from both single crystal and powder XRD data, the O2–O2 distance is short (2.44 Å) as opposed to the ionic radius of oxygen which is 1.36 Å when three-fold coordinated.[36] As the temperature rises up to 550°C, the O2–O2 distance shrinks to approximately 2.35 Å.

At 550°C, the symmetry breaking due to the Cmcm → C2/m structural transition allows the O2 atoms to shift from the 8f position on the mirror plane located at x=0 and x=1/2. It provides to O2 one new degree of freedom to move along the a-axis. The O2 position in the Cmcm space group splits into two 4i positions (i.e O2a and O2b) in the C2/m space group by the absence of the c-glide plane in the space group. At 575°C, the O2a – O2a and O2b – O2b interatomic distances are recovered (i.e. 2.43 Å and 2.44 Å respectively). However, the evolution of the O2a – O2a and O2b – O2b interatomic distances versus temperature between 550°C and the decomposition temperature is different. While the O2a–O2a distance increases up to 2.62 Å, the O2b – O2b distance remains close to 2.34 Å as before the Cmcm → C2/m transition (Fig. 4e). After the first phase transition and the decomposition temperature, M2 site is split. The M2a-M2a and M2b-M2b distances behave differently. The temperature has limited effect on the M2b-M2b which is kept close to 2.79 Å while the M2a-M2a distance increases to 2.9 Å.

In conclusion, the M2b-O2a-M2b-O2a unit in the monoclinic Al$_{1.75}$Ti$_{1.25}$O$_5$ structure, highlighted in green in the Fig. 4f, sees the M2b-M2b unchanged while it is expanded along the O2a-O2a direction. On the other hand, the M2a-O2b-M2a-O2b unit, highlighted in brown, elongates along the M2a-M2a while the O2b-O2b is kept constant.

M containing polyhedron

Heating the $Al_{1.75}Ti_{1.25}O_5$ crystals up to 550°C induce reduced changes to the M1 containing octahedra but modifies the coordination of M2 containing octahedron. The already long apical M2-O1 distance expands further from 2.10 Å to 2.20 Å while the second apical M2-O2 distance shrinks from 1.92 Å to 1.79 Å. Therefore, the coordination of the M2 atoms is gradually changing from a 6-fold distorted octahedron (at room temperature) to a 5-fold distorted square pyramid (at 550°C) as illustrated in Fig. 3b.

After the Cmcm → C2/m phase transition, the M2 as well as the O3 position is split into two, same as O2 into O2a and O2b. At 575°C, both M2a and M2b are 5-fold coordinated. M2a square pyramids are sharing edges and are arranged in layer perpendicular to the $c_{mono}$-axis. The double chain of edge sharing M2b containing square pyramids are running parallel to the $b_{mono}$-axis (Fig. S2). As temperature rises above the C2/m → C2 phase transition, many M-O distances become longer (close to 2.1 Å or higher) and O-M-O angles are modified significantly. At 725°C, only 4 M1-O distances remain below 2 Å with O-M-O getting close to the theoretical angle of a tetrahedron. Similar changes can be observed for M2a indicating that the coordination of M1 and M2a is tending to 4. Above the C2/m → C2 phase transition, the layers built of M2a containing square pyramids are progressively converted into (M2a)2-O4 blocks sharing corner and forming chains running along $b_{mono}$ (Fig. S3). In the case of M2b, only 2 M-O distances remain below 2 Å. However, the polyhedron angles remain close to the ones of a square pyramid.

M site occupancy

As can be seen in Fig. 4c, the Al and Ti distribution in the orthorhombic lattice of $Al_{1.75}Ti_{1.25}O_5$ between the M1 and M2 sites is not equal. M2 is mainly occupied with Al. Taking into account the site occupancy, the formula of the orthorhombic $Al_{1.75}Ti_{1.25}O_5$ can be written as $(Al_{0.52}Ti_{0.48})(Al_{0.62}Ti_{0.38})_2O_5$ using the $M_1(M_2)_2O_5$ formalism. After the orthorhombic to monoclinic transition and the M2 site splitting, M2a occupancy by Al is continuously increasing up to the second phase transition. In the structure with C2 space group, M2a is exclusively occupied by Al. The formula of $Al_{1.75}Ti_{1.25}O_5$ in the C2 symmetry is therefore $(Al_{0.53}Ti_{0.47})(Al_{1.0})(Al_{0.23}Ti_{0.77})O_5$ with $M_1M_{2a}M_{2b}O_5$ formalism. The M2a sites exclusively occupied by Al is suggests a chemical segregation in the structure which could be the initiator of the thermal decomposition of $Al_{1.75}Ti_{1.25}O_5$ into $Al_2O_3$ and $TiO_2$.

**Single crystal quenching**

$Al_{1.75}Ti_{1.25}O_5$ single crystals were annealed at 700°C in vacuum for 12h and subsequently quenched to room temperature. As confirmed by X-ray diffraction, the structure of $Al_{1.75}Ti_{1.25}O_5$ after quenching is orthorhombic Cmcm pseudobrookite. The interatomic distances and the coordination of the M atoms are fully recovered. Therefore, both phase transitions are reversible. It has to be noted that the recovery of the site occupancies was expected. At 700°C, M2a is fully occupied with Al ($occ_{M2a(Al)}=1$). Occupancy of M2b with Al $occ_{M2b(Al)}$ is 0.317. After quenching, M2a and M2b combine as M2 with $occ_{M2(Al)} = 0.606$. Therefore $2\ occ_{M2(Al)} \approx occ_{M2a(Al)} + occ_{M2b(Al)}$ which is consistent with the M2, M2a and M2b site multiplicity being respectively 8f, 4c and 4c.

In summary, $Al_{1.75}Ti_{1.25}O_5$ undergo two reversible phase transitions from the Cmcm pseudobrookite structure to the C2/m monoclinic structure at 550°C and subsequently to C2 monoclinic structure at 650°C. The phase transition is caused by the decrease of the interatomic distance between the two oxygen atoms located in the center of the lattice despite the fact that the lattice expands. The coordination of M is progressively reduced which weakens the structure. This effect associated with a segregation of Al in specific crystal sites leads to the decomposition of $Al_{1.75}Ti_{1.25}O_5$ into $Al_2O_3$ and $TiO_2$ starting from 725°C.

**Physical properties**

Magnetic susceptibility temperature dependence is presented in Fig. 5a. Paramagnetic behavior at high temperatures shows a Curie-Wiess-like behavior. The temperature dependent magnetic susceptibility is fitted, in the temperature range ≈110-320 K, to a Curie-Weiss law,

$$\chi(T) = \chi_0 + \frac{C}{T-\theta} \quad (1)$$

where $C$ is the Curie constant and $\theta$ the Curie-Weiss temperature. Because aluminum has no unpaired electrons in its 3+ valence state, same as titanium in 4+ state, the magnetic moment is assumed to be only on $Ti^{3+}$, $d^1$ atoms (0.25 per unit cell). The best fit to the experimental data (see inset to Figure 5a) yields $\theta$ = -14.8 K, $\chi_0$ = 5.21 $10^{-4}$ emu mol$^{-1}$ Oe$^{-1}$ and C = 0.2618 emu mol$^{-1}$ Oe$^{-1}$ K$^{-1}$ corresponding to an effective magnetic moment of 1.45 $\mu_B$. Calculations assuming spin-only contribution (spin quantum number S = ½, g factor of 3d electron g = 2) give the value of 1.73 $\mu_B$ showing that $Ti^{3+}$ ions appears primarily as free spins. However, the observed effective moment is at the lower range of values observed for $Ti^{3+}$ compounds. Similar behavior of lower number of $Ti^{3+}$ ions obtained when compared to a number calculated from the formal valence was observed in other mixed valence pseudobookite compounds of $Al_{2-x}Ti_{1+x}O_5$ solid solution series (1≤ x ≤ 1.5).[22] Proposed pairing of magnetic ions into the antiferromagnetic spin singlet dimers $Ti^{3+}$- $Ti^{3+}$ could not be inferred from the magnetic susceptibility behavior for this composition. This effect, however, should be limited due to the comparatively low concentration of $Ti^{3+}$ ions. Small contribution could be hidden, together with the effects of crystal defects and impurities from alumina crucible, in the temperature independent susceptibility $\chi_0$, lowering the calculated effective magnetic moment.

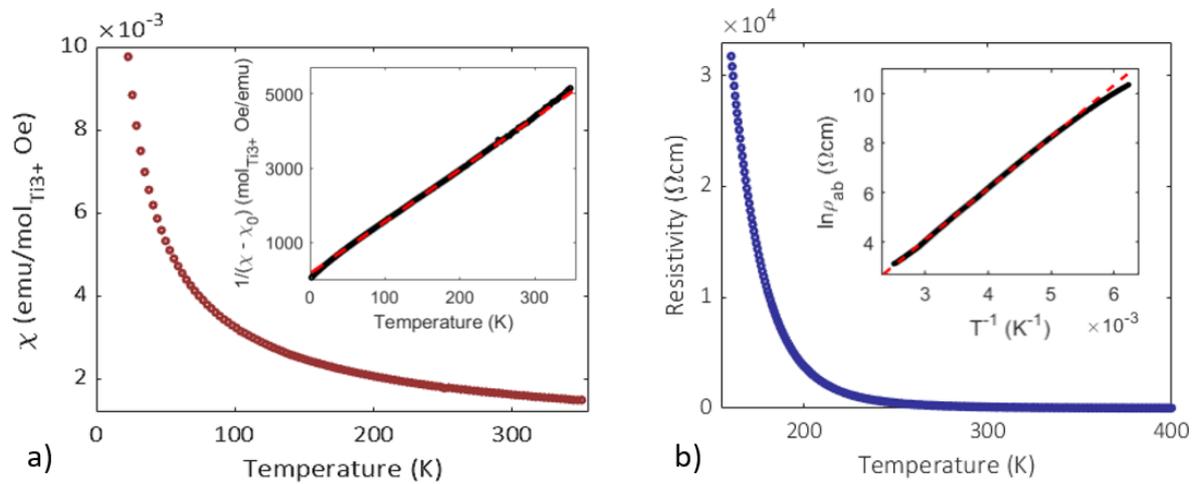

**Fig. 5** a) Temperature dependence of the magnetic susceptibility χ with H = 0.1 T. The inset shows the fitted result using the Curie–Weiss law, where the red line represents the fitting of the measurement. b) Temperature dependence of the resistivity of $Al_{1.75}Ti_{1.25}O_5$ single crystal. The inset shows the plot ln$\rho$ vs $T^{-1}$ to extract the activation energy for conduction.

Fig. 5b shows the temperature dependence of the electrical resistivity for the $Al_{1.75}Ti_{1.25}O_5$ single crystal. It shows semiconducting behavior in the accessible temperature range of 160 to 400 K. Below 160 K, the resistance was too large to be measured. Linear part of the resistivity (190-340 K) was fitted to the thermal activation model using the equation

$$\rho_{ab} = \rho_0 \exp(\frac{E_a}{k_B T}) \quad (2)$$

where $\rho_0$ is the prefactor, $k_B$ is the Boltzmann constant and $E_a$ is the thermal activation energy. The observed room temperature resistivity (135 $\Omega$cm) and activation energy ($E_a$ = 0.18 eV) follow the trend of values reported for other mixed valence aluminum titanate compound with an orthorhombic structure in the solid solution $Al_{2-x}Ti_{1+x}O_5$ (1 ≤ x ≤ 1.5).[22]

The introduction of the magnetic titanium $3d^1$ metal centers significantly alters magnetic and electronic properties of aluminum titanate which can be observed in magnetic susceptibility and electrical resistivity measurements. Nominal aluminum titanate $Al_2TiO_5$ is white while $Al_{1.75}Ti_{1.25}O_5$ is dark blue. Color change can be attributed to a d-d transitions related to the presence of $Ti^{3+}$ ions and a significant decrease in a band gap energy (~3 eV for parent compound $Al_2TiO_5$) as indicated by resistivity measurement.

Considering that aluminum has an oxidation state of +3 and oxygen an oxidation state of -2, the final formula is $Al_{1.75}^{3+}Ti_{0.25}^{3+}Ti_{1.0}^{4+}O_5$. The precursor's EDX compositional analysis shows a compound with a formula $K_{1.7}Ti_2(PO_4)_3$, exhibiting the $Ti^{4+}/Ti^{3+}$ ratio in the same range as in a resulting aluminum titanate. Since it is possible to synthesize series of langbeinite precursors with varying potassium content ($K_{2-x}Ti_2(PO_4)_3$ with 0 ≤ x ≤ 1), precise control over titanium valence state in the final product should be achievable by managing the starting precursor.[24]

## Conclusion

Single crystals of $Al_{1.75}Ti_{1.25}O_5$ were grown by reaction of a $K_{1.7}Ti_2(PO_4)_3$ melt with $Al_2O_3$ at low temperature. Facile method using the alumina crucible as a precursor allows for fast growth of aluminum titanate single crystals with pseudobrookite structure. They undergo a first phase transition from the Cmcm orthorhombic to the C2/m monoclinic symmetry at 550°C and a second transition to the C2 symmetry at 650°C. The temperature-driven evolution of the structure reveals the O-O interatomic distance decrease, despite the lattice expansion, to be the cause of the phase transitions. Furthermore, reduced coordination of Al and Ti atoms at higher temperature is present, together with non-random metal cation distribution. This leads to the instability of $Al_{1.75}Ti_{1.25}O_5$ above 725°C yielding to the thermal decomposition into $Al_2O_3$ and $TiO_2$.

Calculated activation energy and size of magnetic moment for the $Al_{1.75}Ti_{1.25}O_5$ follow the trend in values reported for other mixed valence aluminium titanate compounds, filling the gap in less explored part in solid solution series $Al_{2-x}Ti_{1+x}O_5$. Introduction of even small amount of magnetic titanium $Ti^{3+}$ in parent aluminium titanate ($Al_2TiO_5$) significantly alters its electronic and magnetic properties. Precise control over titanium oxidation state opens up various possibilities for titanium containing pseudobrookite compounds with new structural and electronic properties that show promise in scientific research and industrial application.

**Supplementary information:**
The powder XRD patterns as well as the CIF files of the structure refined at low and high temperature are provided. The crystallographic table for single crystal X-ray diffraction is included. The reciprocal space reconstruction from single crystal XRD data is provided. Rietveld refinement pattern at 25°C, 550°C, 650°C and 725°C are shown.

**Author contribution:**
The single crystals were grown by D.T. The single crystal X-ray diffraction, the high-temperature XRD measurements were performed by W-.H.B. A.M. and D.T. determined the chemical composition of the crystals by the EDX and XRF. W.H.B refined the structure at low temperature from single crystal data. The high temperature structures were refined by the Rietveld method by A.M. D.T. measured the magnetic properties under the supervision of I.Z and H.M.R. Y.L. and D.T. measured the electrical conductivity. D.T., W-.H.B, Y.L., I.Z., H.M.R and A.M analyzed the data. All authors commented on the manuscript.

# SUPPLEMENTARY INFORMATION

**Table S1.** The crystallographic table for single crystal X-ray diffraction at room temperature.

| | |
|---|---|
| Empirical formula | $Al_{1.75}Ti_{1.25}O_5$ |
| Formula weight | 187.05 |
| Temperature (K) | 300.00(2) K |
| Wavelength (Å) | 0.71073 |
| Crystal system | Orthorhombic |
| Space group | *Cmcm* |
| *a* | 3.6353(1) |
| *b* | 9.5509(3) |
| *c* | 9.7391(3) |
| Volume (Å$^3$) | 338.25(2) |
| Z | 4 |
| Calculated density (Mg/m$^3$) | 3.672 |
| Absorption coefficient (mm$^{-1}$) | 4.400 |
| F(000) | 361 |
| Crystal size (mm) | 0.11 × 0.09 × 0.08 |
| Reflections collected/unique | 2866 / 477 [R(int) = 0.0166] |
| Completeness (%) | 100.0 |
| Data/restraints/parameters | 475 / 4 / 32 |
| GOOF | 1.000 |
| $R_1$ [I>2sigma(I)] | 0.0154 |
| $wR_2$ [I>2sigma(I)] | 0.0372 |
| $R_1$ (all data) | 0.0168 |
| $wR_2$ (all data) | 0.0382 |
| Extinction coefficient | 0.0024(8) |
| Largest peak & hole (e. Å$^{-3}$) | 0.636 and -0.807 |

**Figure S1**

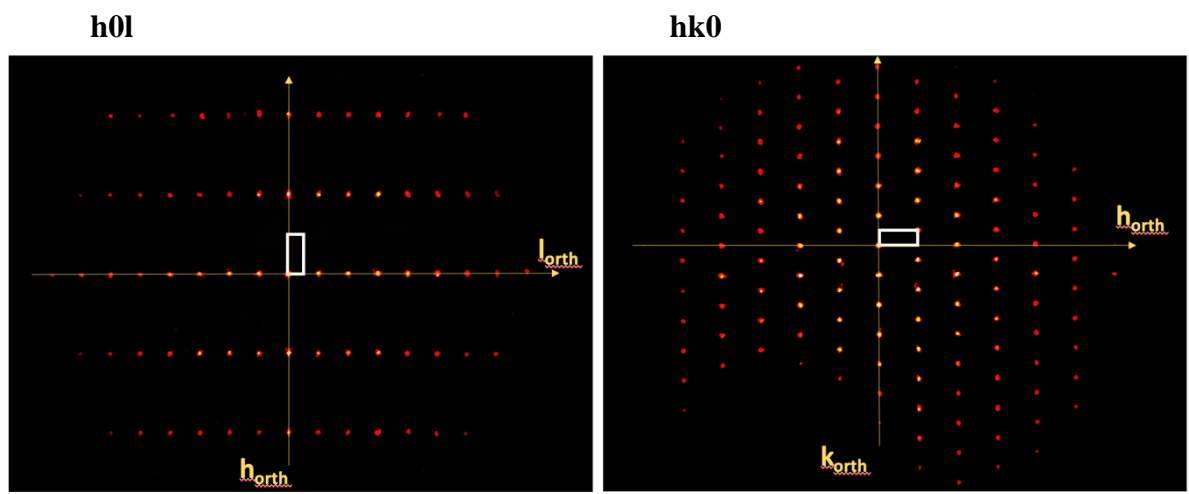

Figure S1: Representative cross sections of the reciprocal space reconstruction for $Al_{1.75}Ti_{1.25}O_5$ crystal. The axes $h_{orth}$, $k_{orth}$ and $l_{orth}$ (in yellow) correspond to the orthorhombic unit cell. The reciprocal unit cell is marked in white.

The obtained systematic extinction of the reflections (hkl: h+k=2n) indicates C centering of the orthorhombic lattice. Second, the systematic extinction of the reflections (h0l: l=2n) indicates the presence of a c-glide mirror perpendicular to the b axis. These observations confirm that the $Al_{1.75}Ti_{1.25}O_5$ crystal correspond to the Cmcm orthorhombic structure of pseudobrookite with lattice parameters $a_{orth}$= 3.63A, $b_{orth}$= 9.55A and $c_{orth}$= 9.73A.

**Figure S2**

**Al$_{1.75}$Ti$_{1.25}$O$_5$ crystal structure at 600°C**

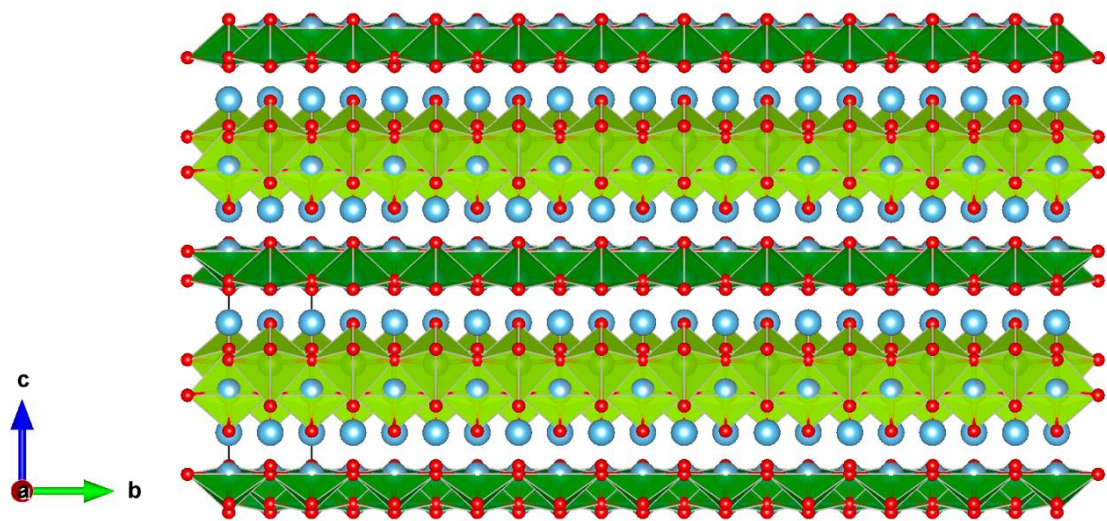

**Al$_{1.75}$Ti$_{1.25}$O$_5$ crystal structure at 725°C**

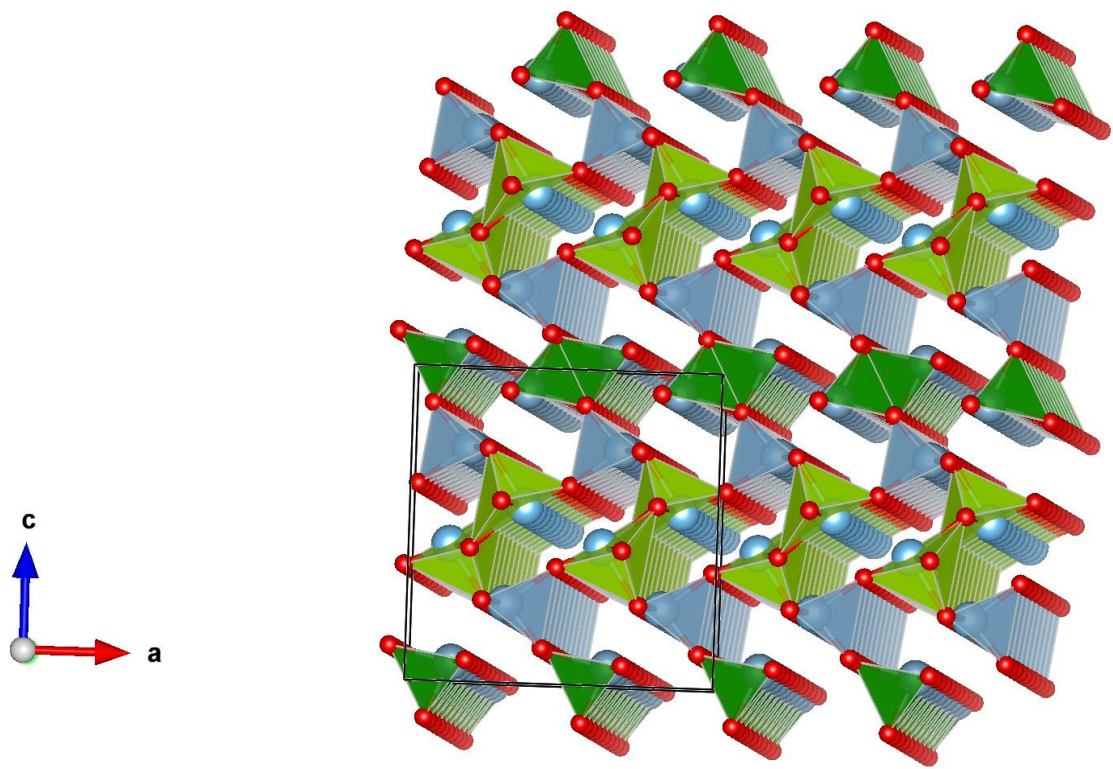

Figure S2. The dark green polyhedrons contain M2a atoms while the light green polyhedrons contain M2b atoms.

Rietveld refinement results
Al$_{1.75}$Ti$_{1.25}$O$_5$ @ 25°C

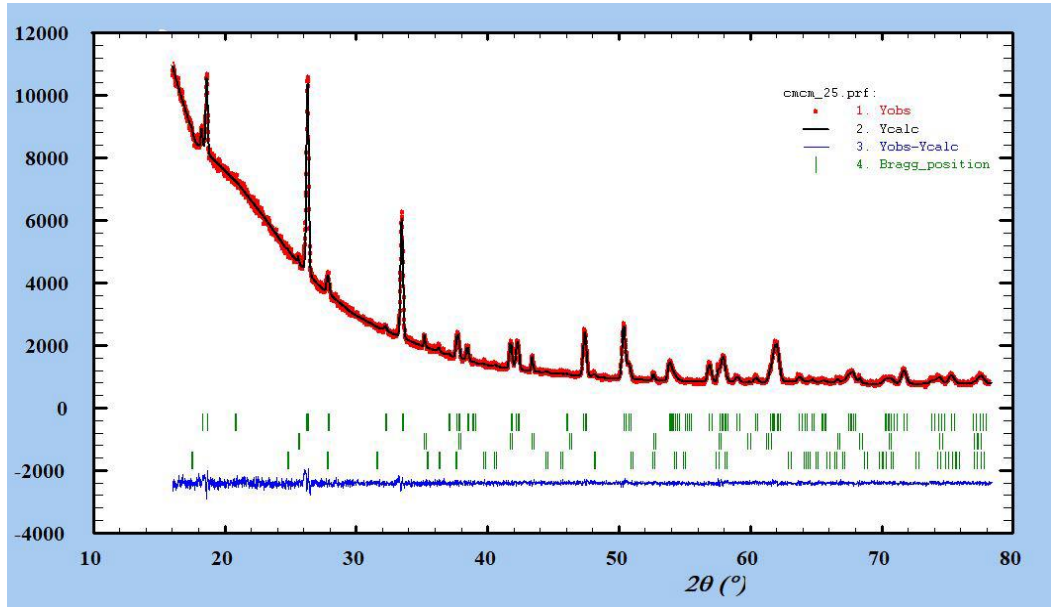

Al$_{1.75}$Ti$_{1.25}$O$_5$ @ 550°C

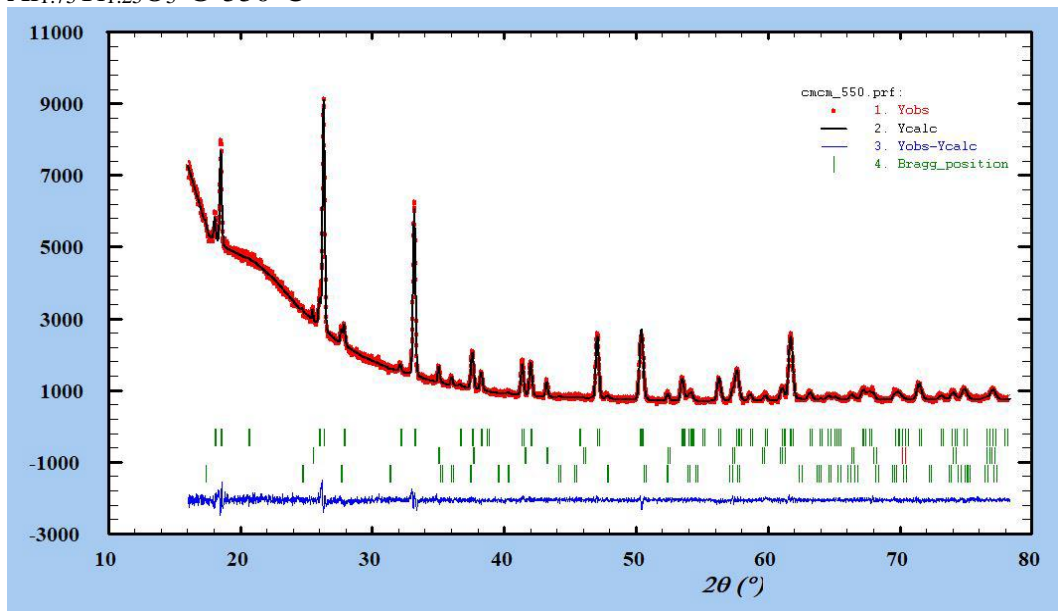

Al$_{1.75}$Ti$_{1.25}$O$_5$ @ 550°C

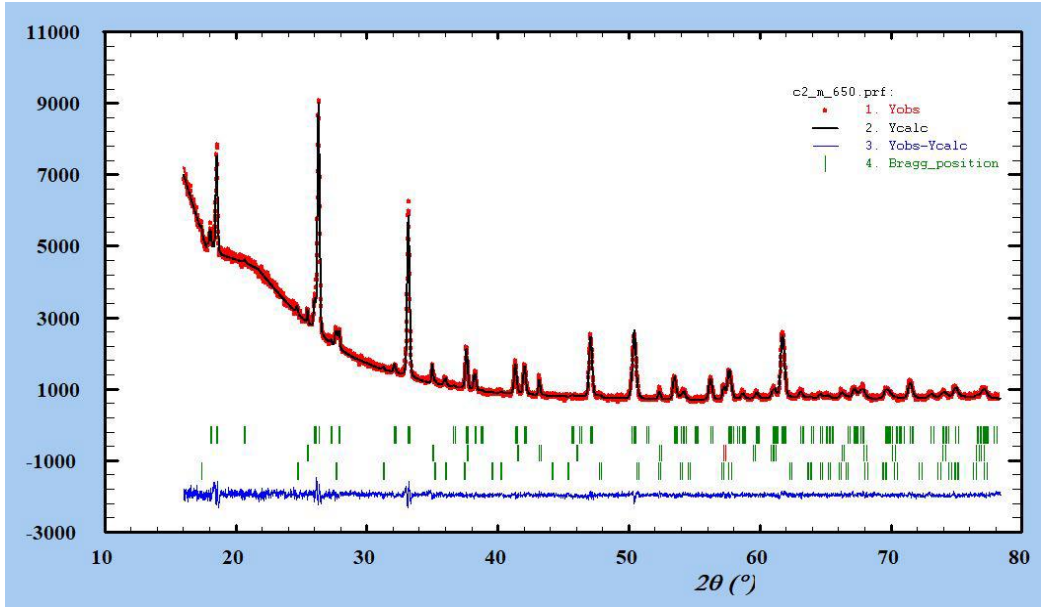

Al$_{1.75}$Ti$_{1.25}$O$_5$ @ 725°C

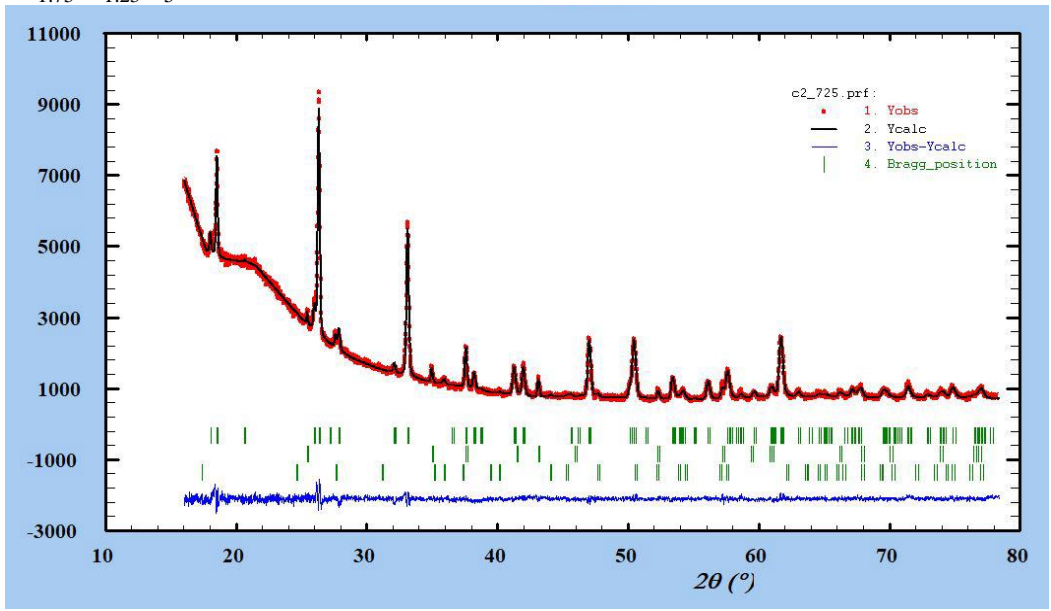